\documentclass[aps,preprint,amsmath,amssymb]{revtex4}
\usepackage{amssymb}
\usepackage{mathrsfs}
\usepackage{graphicx}
\usepackage{subfigure}
\usepackage{hyperref}
\usepackage{dcolumn}
\usepackage{bm}
\usepackage{txfonts}
\usepackage{booktabs}
\usepackage{multirow}

\newcommand{\mycite}[1]{\scalebox{1.0}[1.0]{\raisebox{0.1ex}{\cite{#1}}}}
\usepackage[labelsep=period]{caption}
\usepackage{hyperref}
\begin{document}
\title{CE Screen: an energy-based structure selection method}
\author{Zongguo Wang}
\affiliation{Computer Network Information Center,
Chinese Academy of Science, Beijing 100190, People's Republic of China
} \email{wangzg@cnic.cn}
\author{Xiaoyu Yang}
\affiliation{Computer Network Information Center,
Chinese Academy of Science, Beijing 100190, People's Republic of China}
\author{Ligen Wang}
\affiliation{General Research Institute for Nonferrous Metals,
Beijing 100088, People's Republic of China}
\author{Juan Wang}
\affiliation{Computer Network Information Center,
Chinese Academy of Science, Beijing 100190, People's Republic of China}
\affiliation{University of Chinese Academy of Sciences, Beijing, 100049, China}
\author{Xushan Zhao}
\affiliation{Computer Network Information Center,
Chinese Academy of Science, Beijing 100190, People's Republic of China}
\author{Mingming Zhang}
\affiliation{Computer Network Information Center,
Chinese Academy of Science, Beijing 100190, People's Republic of China}
\affiliation{University of Chinese Academy of Sciences, Beijing, 100049, China}
\author{Jie Ren}
\affiliation{Computer Network Information Center,
Chinese Academy of Science, Beijing 100190, People's Republic of China}
\affiliation{University of Chinese Academy of Sciences, Beijing, 100049, China}

\begin{abstract}
We have developed a method to improve the doping computation efficiency, this method is based on first principles calculations
and cluster expansion. First principles codes produce highly accurate total energies and optimized geometries for any given structure.
Cluster expansion method constructs a cluster expansion using partial first principles results and
computes the energies for other structures derived from a parent lattice.
Using this method, energies for multiple doping structures can be predicted quickly without series of first principles calculations.
This method has been packaged into a tool named as CE Screen and integrated into MatCloud (A high-throughput first principles calculation platform).
This makes the tool simple and easy for all the users.

\end{abstract}
\maketitle
\section{Introduction}
More and more materials problems are made approachable by computation simulation
with the growing of computer power and improvements of calculation methods.
Since the last century, first principles simulations are powerful tools, because they allow for
the exploration of new materials before an experiment devises them. Doping, as an effective way
to improve the performance of known materials, is also considered as an important way to construct
structures for new materials.
In order to obtain the stable structures from doping structures with all different concentrations,
a large number first principles calculations for possible configurations are needed to be carried out.
The high-throughput(HT) method, which involves setting up and performing ab initio calculations,
reorganizing and analyzing the results with minimal intervention by the user,
has become an effective and efficient tool for materials development and prediction\cite{Setyawan2010}.

Since the Materials Genome Initiative(MGI) was launched in 2011 in the United States\cite{KalilTomAndWadia2011},
many high-throughput calculation platforms and codes have been developed, such as,
a software framework named Automatic Flow (AFLOW)\cite{Curtarolo2012}for
HT calculation of crystal structures developed by Curtarolo group,
a core program named Materials Project\cite{Jain2013} developed by Ceder group,
and a platform named MatCloud(\url{http://matcloud.cnic.cn}) developed by Yang group.
Although these HT method and their implementation accelerate the process of materials exploration,
some sort of automatic optimization technique are still needed to develop and integrated in these HT platforms.
As we known, finding the most stable crystal structures of compounds is one of the classical problems in
inorganic materials, because knowing the stable structure holds the key to material properties.
While direct quantum mechanical calculations for all doping configurations in search of the most
stable structure is not computationally feasible,
even using crystallographic equivalence or on the limited number of occupying sites, it is still a tedious problem.
Therefore, to improve the efficiency of doping calculations,
a method to quickly screen stable structures from enormous doping structures should be developed.

Since cluster expansion was proposed by Mayer in 1941, it has been employed as an approximate
computation method to express the partition function as power series expansion\cite{Wu2016}.
Cluster expansion (CE) method has been a very compact and efficiency way to represent the alloy energetics\cite{Chakraborty2010},
and it can be constructed using the alloy theoretic automated toolkit(ATAT)\cite{VandeWalle2002}.
CE method has recently been invoked to explore the stability of various two-dimensional and three-dimensional materials
except for alloys\cite{Ravi2010,Penev2012,Li2014,Kutana2014}.
These applications have approved that CE is a good method to predict new structures and their energies.
However, existing methods based on CE are unable to handle the special situations that selecting stable structures from all the known configurations.
This will give rise to the inconvenience of researchers.

In this work, we will introduce a method to screen stable structure.
It can quickly screen the stable structures from
a large number of known structures with less first-principle calculations.
This method is also based on cluster expansion(CE) method which is a classical multiscale model.
In additions, this method has been packaged into a tool named CE Screen and integrated in the HT platform MatCloud.
People also can get this tool from the address of MatCloud (\url{http://matcloud.cnic.cn}).
The organization of this paper is as follows,
first, in theoretical aspect, we briefly introduce supercell approach, CE method, and
parameters of the first principles calculations.
Second, we depict how the CE Screen works,
and the application method of this tool on MatCloud platform is also supplied.
Finally, we apply CE Screen for two systems, which are fcc Al-Ti and bcc Fe-Al, these two cases
doping Ti into fcc Al and doping Al into bcc Fe serve as the test
cases to evaluate the implementation with respect to literature data.
\section{Theories and Methods}
The process of CE Screen in doping system is in three steps.
In the first step, generating all the doped structures and selecting
some structures (recommended structure number is $30\sim50$)for energy calculation,
the number of fitting structures will be determined by the practical cases.
In this step, supercell approach and crystallographic equivalence method are used.
The second step, the effective cluster interactions (ECIs) should be determined using CE method.
The last step, predicting energies for other doped structures,
and then recommending few structures for each concentration to do first principles calculation,
$3\sim8$ structures are recommended for each concentration except that concentrations are $0$ and $1$.
This section has three parts. Firstly, we briefly review the supercell construction method,
secondly, we describe the CE method and propose the calculation method of ECIs,
Finally, the first principles calculation parameters are supplied for our test cases.
\subsection{supercell construction}
Point defects consist in atomic substitutions or vacancies, meaning that the nature of the atoms
occupying one or several crystallographic sites is changed.
A supercell is essential for studying lattice vibrations
and to build up structures with substitutional, interstitial or magnetic disorder. Frequently,
structure compositions supplied by experiments are partial occupancies.
In order to describe the structure as a regular periodic structure with well-defined parameters,
a spacegroup and a set of crystallographic sites, and each crystallographic
site is strictly occupied with a single type of atom, a supercell is needed. Within this supercell,
numerous atomic configurations can be compatible with the partial occupancies.
A supercell is a repeating unit cell of the crystal that contains several primitive cells.
The lattice vectors of unit cell is $(\vec{a},\vec{b},\vec{c})^{T}$,
a $3\times3$ coefficient matrix $A$ is the multiples for each direction:
\begin{equation}\label{supercell}
  \left(
    \begin{array}{c}
      \vec{a}^{'} \\
      \vec{b}^{'} \\
      \vec{c}^{'} \\
    \end{array}
  \right)=\left(
            \begin{array}{ccc}
              A_{11} & A_{12} & A_{13} \\
              A_{21} & A_{22} & A_{23} \\
              A_{31} & A_{32} & A_{33} \\
            \end{array}
          \right)*\left(
                    \begin{array}{c}
                      \vec{a} \\
                      \vec{b} \\
                      \vec{c} \\
                    \end{array}
                  \right)
\end{equation}
Given coefficient matrix $A$, we can obtain any multiples of unit cell.
At the same time, all the atomic sites should be multiplied according to
invariance of translation.
\subsection{Cluster expansion}
We briefly describe the CE method following Refs.\mycite{Sanchez1984,Wolverton1993,Wolverton1994,Zunger2002}.
The configurations of a crystalline system with n lattice sites is described by
characterizing each lattice site $p$ a spin or occupation variable $\sigma_{p}$.
$\sigma_{p}$ can take the values $\pm m$,$\pm(m-1)$,$...$,$\pm1$, and
0 for an $M-$ component system where $M=2m$ (or $2m+1$).
Any configuration of the spin variables is fully specified by
the $n-$ dimensional vector $\vec{\sigma}=({\sigma_{1},\sigma_{2},...,\sigma_{n}})$.
In the doping case, i.e. atomic substitutions or vacancies,
the spin variable take the value $\sigma_{p}=\pm1$, which depending on the type of atom
occupying the site.

For a given cluster of a set of lattice sites $\alpha={1,2,...,n}$
and a set of functional indices $s={n_{1},n_{2},..n_{n}}$, the cluster function $\Phi_{\alpha}^{s}$ defined as
\begin{equation}\label{ce-function}
  \Phi_{\alpha}^{s}(\vec{\sigma})=\Theta_{n_{1}}\sigma_{1}\Theta_{n_{2}}\sigma_{2}...\Theta_{n_{n}}\sigma_{n}=\prod_{i\in\alpha}\sigma_{i}
\end{equation}
It is a characteristic function of the cluster $\alpha$ defined by the lattice sites ${1,2,...,n}$.
These cluster functions form a completeness orthonormal basis, the scalar product between two arbitrary functions is
\begin{equation}\label{scalar-product}
  \langle\Phi_{\alpha},\Phi_{\beta}\rangle=\delta_{\alpha,\beta}
\end{equation}
We can express any cluster function of the configuration $\vec{\sigma}$
with the set $\Phi_{\alpha}$ in terms of the Chebychev cluster functions of Eq.\ref{ce-function}\cite{Wolverton1994} as
\begin{equation}\label{function}
  F(\vec{\sigma})=\sum_{\alpha}F_{\alpha}\Phi_{\alpha}(\vec{\sigma})=F_{0}+\sum_{\alpha}F_{\alpha}\prod_{i\in\alpha}\sigma_{i}
\end{equation}
The summation takes over all the clusters $\alpha$ in the crystal,$F_{0}$ is the configuration-independent term,
the components $F_{\alpha}$ is as the expansion coefficients
\begin{equation}\label{coefficients}
  F_{\alpha}=\left\langle\prod_{i\in\alpha}\sigma_{i},F(\vec{\sigma})\right\rangle
\end{equation}
The configurational energy may be expanded in cluster expansions:
\begin{equation}\label{configuration-energy}
  E(\vec{\sigma})=V_{0}+\sum_{\alpha}V_{\alpha}\sigma_{\alpha}
\end{equation}
The formation energy of a particular configuration,$E$, is expanded exactly in polynomials of the spin variables weighted by
multisite interaction parameters {J} as an Ising-like form\cite{Herder2015}
\begin{equation}\label{energy-expanded}
  \Delta E(\vec{\sigma})=J_{0}+\sum_{i}^{sites}J_{i}\sigma_{i}+\sum_{ij}^{pairs}J_{ij}\sigma_{i}\sigma_{j}+\sum_{ijk}^{pairs}J_{ijk}\sigma_{i}\sigma_{j}\sigma_{k}+...
\end{equation}
It can approximately expressed as a series expansion of "clusters" according to Eq.\ref{configuration-energy}:
\begin{equation}\label{CE}
  \Delta E_{CE}(\vec{\sigma})=\sum_{\alpha}m_{\alpha}J_{\alpha}\left\langle\prod_{i\in\alpha^{'}}\sigma_{i}\right\rangle
\end{equation}
Where $\alpha$ is a cluster (a set of sites $i$). The summation takes over all clusters $\alpha$ that not equivalent
by a symmetry operation of the space group of parent lattice,
the average is taken over all the clusters $\alpha^{'}$ that are equivalent to $\alpha$
by symmetry. The coefficients $J_{\alpha}$ in this expansion embody the information regarding
the energetics of the system and are called the effective
cluster interaction (ECI). The multiplicities $m_{\alpha}$ indicate the number of clusters that are equivalent to $\alpha$.

Combining the first and the third product factors of Eq.\ref{CE}, it can be expressed as
\begin{equation}\label{CE-formulated}
  \Delta E_{CE}(\vec{\sigma})=\sum_{f}J_{f}\overline{\prod_{f}}(\vec{\sigma})
\end{equation}
Where, $\prod_{f}(\vec{\sigma})$ are the spin-products averaged over the entire lattice and they are formulated per lattice site.
$f$ are the all possible types of clusters found among the lattice sites.
From Eq.\ref{CE-formulated}, finding the ECIs($J$) is a simple linear algebra problem.
For each configuration $\vec{\sigma}$, we have an equation
with a unique value of E, unique values for the $\prod_{f}(\vec{\sigma})$ and unknown $J$.
This linear equation forms a matrix inverse calculations.
For example, five structures from the same parent lattice are calculated using first principles calculation,
the $\prod$ matrix including cluster information is
computed ($n$ is the number of considered clusters), the ECIs can be found by inversion:
\begin{equation}\label{EJ}
  \left(
    \begin{array}{c}
      E_{1} \\
      E_{2} \\
      E_{3} \\
      E_{4} \\
      E_{5} \\
    \end{array}
  \right)=\left(
            \begin{array}{cccc}
              \prod_{1,1} & \prod_{1,2} & ... & \prod_{1,n} \\
              \prod_{2,1} & \prod_{2,2} & ... & \prod_{2,n} \\
              \prod_{3,1} & \prod_{3,2} & ... & \prod_{3,n} \\
              \prod_{4,1} & \prod_{4,2} & ... & \prod_{4,n} \\
              \prod_{5,1} & \prod_{5,2} & ... & \prod_{5,n} \\
            \end{array}
          \right)\left(
                   \begin{array}{c}
                     J_{1} \\
                     J_{2} \\
                     J_{3} \\
                     J_{4} \\
                     J_{5} \\
                   \end{array}
                 \right)
\end{equation}
Hence,
\begin{equation}\label{EJ-inverse}
   \left(
     \begin{array}{c}
       J_{1} \\
       J_{2} \\
       J_{3} \\
       J_{4} \\
       J_{5} \\
     \end{array}
   \right)=\left(
              \begin{array}{cccc}
                \prod_{1,1} & \prod_{1,2} & ... & \prod_{1,n} \\
                \prod_{2,1} & \prod_{2,2} & ... & \prod_{2,n} \\
                \prod_{3,1} & \prod_{3,2} & ... & \prod_{3,n} \\
                \prod_{4,1} & \prod_{4,2} & ... & \prod_{4,n} \\
                \prod_{5,1} & \prod_{5,2} & ... & \prod_{5,n} \\
              \end{array}
            \right)^{-1}
            \left(
              \begin{array}{c}
                E_{1} \\
                E_{2} \\
                E_{3} \\
                E_{4} \\
                E_{5} \\
              \end{array}
            \right)
\end{equation}
Using the $E$'s given above for each structure, invert $\prod$ matrix that we computed and use it to obtain the ECIs.
$i$ and $j$ in $\prod_{i,j}$ are structure and cluster label,
$j$ is the cluster type number, for the empty cluster, $\prod_{i,1}$ is always 1.
$\prod$ matrix is not always a square matrix, so Least Squares Fit(LSF) need to be used.
\subsection{First principles calculation}\label{parameters}
The energies $E$ of all the fitting structures are calculated using density functional theory by first
principles codes, the first principles code packages including VASP and abinit are supported by MatCloud.
CE screen tool automatically call these calculation code to compute energies of specific structures.
As a default, projector augmented wave (PAW) pseudopotentials with generalized gradient approximation(GGA)
exchange correlation functionals as parameterized by Perdew, Burke and Ernzerhof was employed.
In order to obtain an accurate measure of total energy, each structure is fully relaxed
with a convergence tolerance of 0.1meV/atom using dense grids of
$8000$ $k$-points per reciprocal atom.
\section{CE Screen}
\begin{figure}
  \centering
  \includegraphics[width=\textwidth]{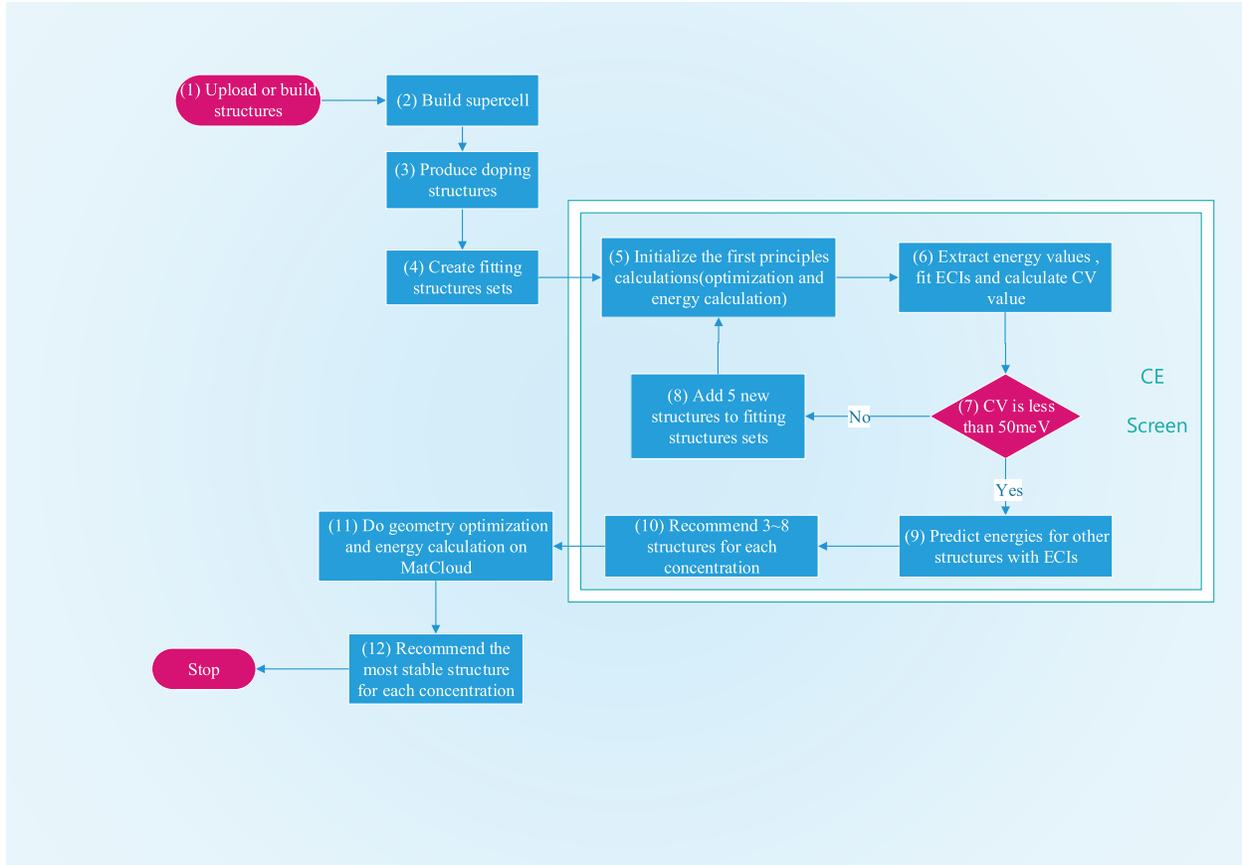}
  \caption{Schematic overview of the cluster expansion (CE) Screen process on MatCloud.}\label{process}
\end{figure}

To find the most stable structures for each concentration in all doping structures,
CE Screen including first principles calculations and cluster expansion should be used.
First principles calculations supply energies to fit the ECIs,
and it also be used to determine the most stable structure for each concentration.
All the processes of selecting doping ground structures on MatCloud are shown in Fig.\ref{process}.
The iterative process should be carried out efficiently to obtain maximum CE predictive reliability
with minimum computational cost. The first principles calculations are the slowest step,
so the algorithms for fitting structures and clusters selection are critical.
In this section, we will describe the structure selection algorithms and depict the work process of this tool on MatCloud in details.
\subsection{Clusters selection}
We also employed the MIT ab initio Phase Stability(MAPS) cluster selection algorithm
that was used in ATAT\cite{VandeWalle2002} to facilitate cluster selection.
It favors compact clusters and requires that all the sub-clusters of
a many-body cluster are included in a CE\cite{Zarkevich2004,Herder2015}.
An input file containing parent lattice information is needed to use CE Screen.
The clusters information are determined according to parent lattice data in the input file and its symmetry operations.
The size of maximal cluster is determined by the radius cutoff specified by user, and we set the default
value is the 8th nearest neighbor of parent structure referring to $N4R8$ criterion\cite{Zhang2016}.
For MatCloud platform, the input file will be automatically generated according to the initial structures uploaded by user.
The formula has three parts,
the first one is coordinate system containing lattice parameters data, the second is lattice vectors for the
unit cell (either a primitive cell or a standard cell is ok),
and the third are the atoms in the lattice, atoms include information of atomic symbols and atomic coordinates.
\subsection{Fitting structures}\label{select_rules}
In order to improve the calculation efficiency, we make some rules for structures in the fitting sets:

(1) The structures selected should cover all the concentrations, i.e.,
at least one structure has been selected per concentration;

(2) All the selected structures should be symmetrically in-equivalent structures.

and (3) The structures should include all different sized supercells if there is.

Structural relaxation and energy calculations are carried out for all structures in the fitting sets,
calculation parameters in section\ref{parameters} are used.
After all the first principles calculations are finished,
the energies of these fitting structures will be extracted and their clusters identified by parent lattice are calculated at the same time.
And then, the ECIs can be obtained through LSF by Eq.\ref{EJ-inverse}.
Using these ECIs, energies of other doping structures will be rapidly predicted using Eq.\ref{EJ}.
In order to assess the predictive power of the this tool, the cross validation(CV) score is used.
It is defined as
\begin{equation}\label{cv-score}
  CV=\left(\frac{1}{n}\sum_{i=1}^{n}(E_{i}-\hat{E}_{i})^{2}\right)^{\frac{1}{2}}
\end{equation}
where $E_{i}$ is the calculated energy of structure $i$, while $\hat{E}_{i}$ is the predicted value of the energy of structure $i$ obtained from
a LSF to the (n-1) other structural energies ($n$ is the fitting structure number).
New calculated structures will be added to be used to fit ECIs
until the CV is less than the tolerance users have set(The maximum CV value default is 50meV).
\subsection{Applications on MatCloud}
The CE Screen has been integrated into MatCloud,
users just need to supply the initial structures which are used to be doped and the doping concentrations users are
interested in, MatCloud will automatically call the CE Screen to
recommend the stable structures for different concentrations.
Users will receive an email when all the process are finished.
Here is a brief outline of steps about CE Screen on MatCloud platform which is implemented in Fig.\ref{process}.

(1)Upload (or build) initial structures in cif(crystallographic information file) or other formations.

Files containing parent lattice information will be generated.

(2)Construct supercells depending on your requirements. In this step, the input is a $3\times3$ matrix.

(3)Produce doping structures.

(4)Create fitting structures sets.

When doped elements and concentrations are identified, MatCloud will produce all the required doping structures, and
select structures by using symmetrically equivalent tool to create fitting structures sets. These structures are used to
fit the ECIs.

(5)Initialize the first principles calculations.

MatCloud will automatically supply the calculated parameters according to the first principles code and tasks,
users can change them if you would. MatCloud submit the jobs of geometry optimization and energy calculation
for all fitting structures to HPC.

(6)$\sim$(8)Obtain the ECIs and its CV value.

All the energies of fitting structure are extracted after first principles calculations are finished,
and then the ECIs will be fitted by using energy and cluster's information of each calculated structure.
Energies of all the fitting structures will be predicted with the ECIs, and the CV value is also computed.
Add new structures to fitting structures sets and iterate steps (4)$\sim$(8) until maximum CE predictive reliability is obtained.

(9)and (10)Predict energies for all doping structures with the well fitted ECIs, and then recommend the stable structures with different
concentrations for user.

After step (10), users can do first principles calculations for these recommended structures to select the lowest energy structure for
each concentration.

Although there are many steps shown in Fig.\ref{process}, time-consuming processes are handled by MatCloud.
Users only need to wait for the e-mail notification of the job status after the information of (1)$\sim$(3) is identified.
(4)$\sim$(10) is the process of CE Screen. It greatly improves the efficiency of selecting stable structures from multiple doping structures.
In additions, MatCloud platform which integrates the supercell and doping structures construction, and EC Screen,
make all the calculations more convenient.
\section{Test cases}
\begin{figure}
\includegraphics[width=\textwidth]{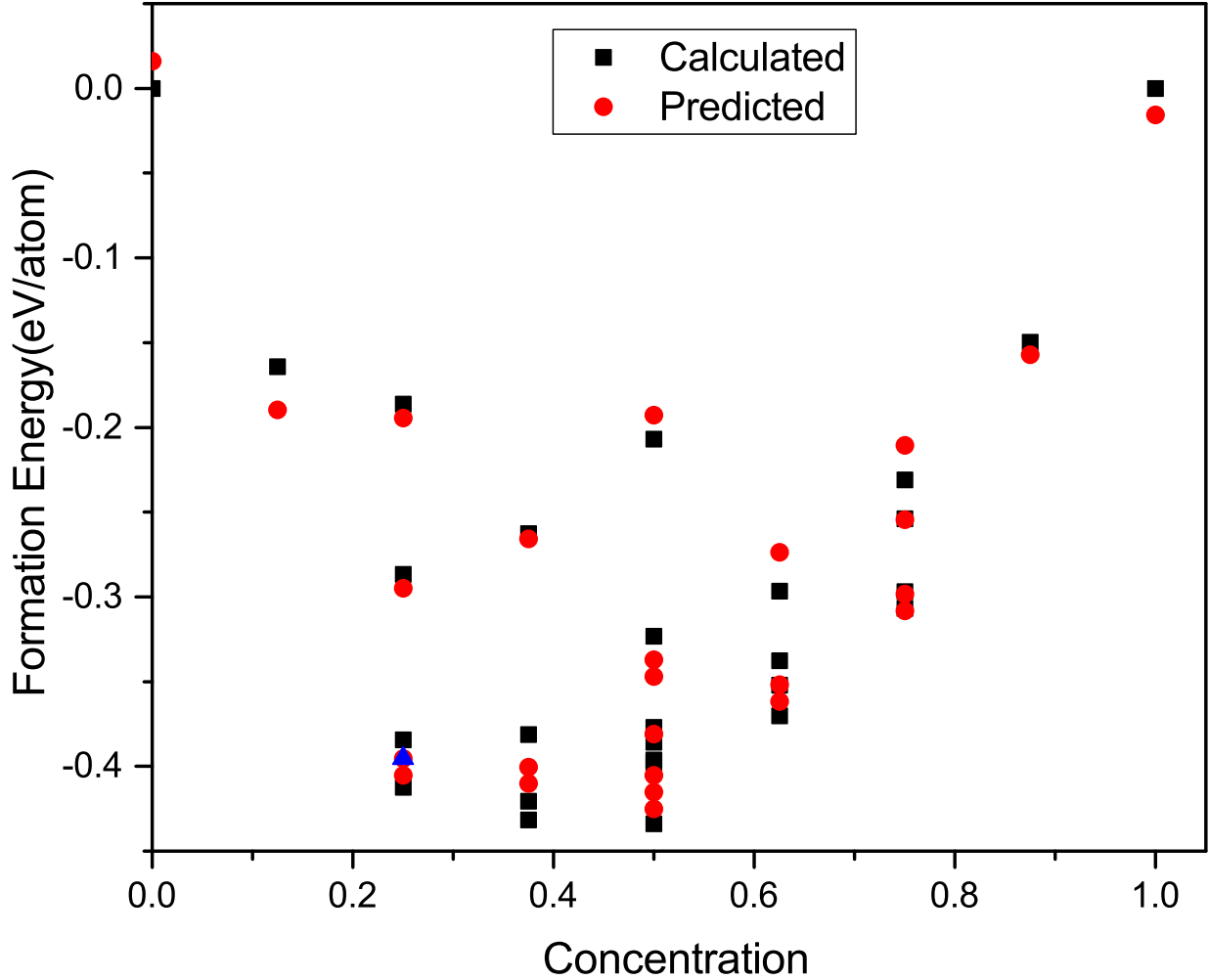}
\caption{Formation energies for 27 $Al_{1-x}Ti_{x}$ structures from first principle calculations and CE Screen tools performed.
Black squares denote energies from first principles, red circles denote energies predicted from CE Screen tools, and the blue upTriangles
emphasize the predicted energies of structures which have been confirmed as the ground states from first principles.}\label{Al-Ti}
\end{figure}

The CE Screen integrated in MatCloud can in principle be applied to any doping systems,
i.e. substitution, vacancy, adsorption and interstitial sites. In this paper, we select the stable structures
with different concentrations from all the doping structures for two systems, fcc Al-Ti and bcc Fe-Al.
Al-Ti and Fe-Al are well-studied, technologically very important systems\cite{Chakraborty2010,Ghosh2008}.
Hence, we choose these two systems as the prototypes to test our tools.
In order to compare the calculated efficiency and make calculations executable, we select small amount structures.
A fcc Al supercell of $2\times1\times1$ with dopant Ti is calculated, nine doping concentrations are considered:
0, $\frac{1}{8}$, $\frac{2}{8}$, $\cdots$ and 1.
The number of all the doping structures is 257, and it is reduced to 27 after operating the crystallographic equivalence to all
these initial doping structures. We calculated energies of all these 27 structures on MatCloud.
We chose 20 doping structures randomly as fitting structures to predict energies for other 7 structures.
The calculated and fitted formation energies for these 27 structures are shown in Fig.\ref{Al-Ti}.
The formation energies for $Al_{1-x}Ti_{x}$ are calculated by
\begin{equation}\label{Al-Ti-E}
  \Delta{E}=\frac{1}{8}E_{doped}-(1-x)E_{Al}-xE_{Ti}
\end{equation}
Where, $E_{system}$ is the total energy for doped structure, $E_{Al}$ and $E_{Ti}$ are the atomic energy for per Al and Ti, respectively.
We use all the 27 structures to fit a set of ECIs, and predict energies for these 27 structures. The CV value for $Al_{1-x}Ti_{x}$ is 0.021eV.
The ground states for each concentration from first principles calculation and CE Screen
tools are similar, except that the concentrations are 0.25. This error can be avoid by adding recommended structures number for each concentration.

\begin{figure}
\includegraphics[width=\textwidth]{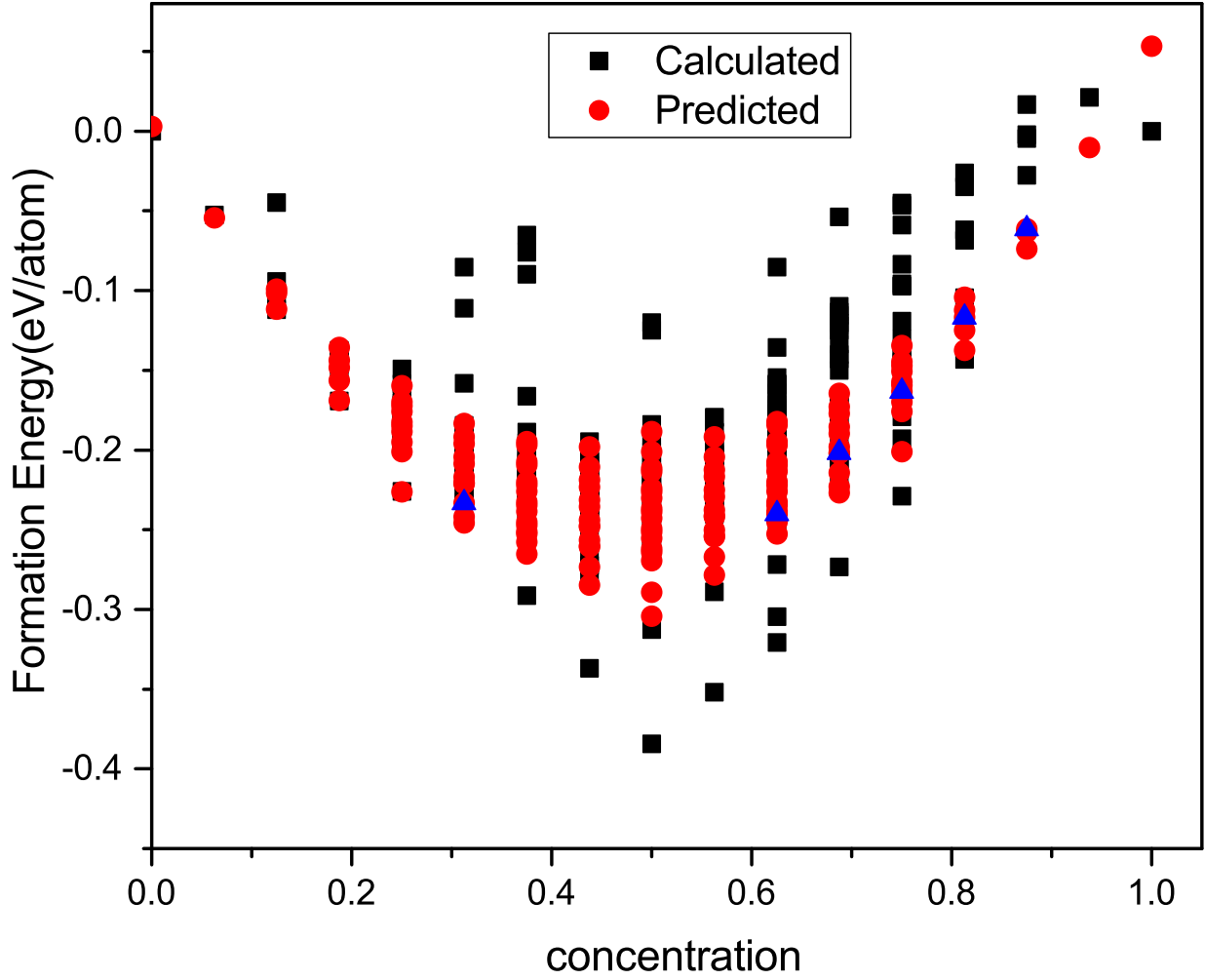}
\caption{Formation energies for 331 $Fe_{1-x}Al_{x}$ structures from first principle calculations and CE Screen tools performed.
Black squares denote energies from first principles, red circles denote energies predicted from CE Screen tools, and the blue
upTriangles emphasize the predicted energies of structures which have been confirmed as the ground states from first principles.}\label{Fe-Al}
\end{figure}

Another example is a two-component Fe-Al system, a bcc Fe supercell of $2\times2\times2$ with dopant Al.
To study the solution doping effects on Fe supercell, seventeen doping concentrations are
considered: 0, $\frac{1}{16}$, $\frac{2}{16}$, $\cdots$, and 1. The number of all the doping structures is 65536, and it is reduced to 331 after
operating the crystallographic equivalence to all these initial doping structures. To select the stable structures for each concentration,
usually we need to calculate all the structures with different concentrations.
CE Screen tools will speed up this process, it only needs at most to calculate 100 structures according to our rules.
We firstly chose 30 doping structures following the rules depicted in section\ref{select_rules} as fitting structures sets.
The calculated CV value for $Fe_{1-x}Al_{x}$ is 0.04eV, satisfying the default tolerance (50meV).
And then use the fitted ECIs and the cluster's information of the rest 301 structures to predict their energies.
The second step is first principles calculations for selected 8 structures from each concentration, and we should make sure that
the number of all structures needed to be calculated is less than 100.
The calculated and predicted formation energies for these 331 structures are
shown in Fig.\ref{Fe-Al}. The formation energies for $Fe_{1-x}Al_{x}$ are calculated by
\begin{equation}\label{Fe-Al-E}
  \Delta{E}=\frac{1}{16}E_{doped}-(1-x)E_{Al}-xE_{Ti}
\end{equation}
Where, $E_{doped}$ is the total energy for doped structure, $E_{Fe}$ and $E_{Al}$ are the atomic energy for per Fe and Al, respectively.
From Fig.\ref{Fe-Al}, although not all the most stable structure are predicted exactly, the ground states for each concentration are all
sit in the candidates. The largest number of structures to be calculated is 5 when concentration is $\frac{12}{16}$. It's worth mentioning that
the most stable structures predicted from CE Screen tool are the same as that calculated from first principles calculations when the concentrations
are 0, $\frac{1}{16}$, $\frac{2}{16}$ , $\frac{3}{16}$, $\frac{4}{16}$, $\frac{6}{16}$, $\frac{7}{16}$,
$\frac{8}{16}$, $\frac{9}{16}$, $\frac{15}{16}$, and $1$.
From the calculated results, we can claim that the ground states for other concentrations
can be well predicted by calculating at most 5 additional doping structures.
\section{Conclusions}
To truly predict the structure of a doping compound, we firstly must find the lowest-energy structure at a concentration.
In this paper, we developed a method to select stable structures for different doping concentrations
with a small number of first principles calculations.
This method has combined approach of first-principles calculations and cluster expansion method, and it
beautifully solves the problem that quickly selecting lowest-energy structures from a huge number of known structures.
This method has been packaged into a tool, named as CE Screen, and it will be download from address of MatCloud (\url{http://matcloud.cnic.cn}) soon.
The tool has been integrated into MatCloud platform which developed by our group, and this makes it simple and easy for all the users.

In application cases, energy calculations of two doping systems have been carried out employing two different methods.
One is completely first principles calculations, the other is using CE Screen.
By using CE Screen on Matcloud, the survey of only 30 symmetrically in-equivalent configurations is sufficient
to find the ground state structure of $Fe_{1-x}Al_{x}$ within the whole range of the concentrations.
Although a worse result for hcp structures has been published, a better-converged result can be achieved
by increasing relatively more input fitting structures.

Although CE Screen developed by our group is only suitable for structures derived from a parent lattice,
it will be a powerful tool for evaluating the stability of these structures.
Two important facts should be pointed if you want to use this tool on MatCloud.
One is that the number of input fitting structures required for the
CE Screen is started from 30, it also varies from system to system.
The other is that first principle codes of Abinit, VASP, CASTEP, PWSCF, $\cdots$, are all supported by CE Screen and
MatCloud, however, users should supply the license if you want to use commercial softwares such as VASP and CASTEP.
\section*{Acknowledgment}
This work was supported by NNSF11547177 of China and the platform MatCloud.
\bibliographystyle{unsrt}
\bibliography{CE}
\end{document}